\documentclass[12pt]{article}
\usepackage{graphicx}
\usepackage{subfigure}
\usepackage{amsmath}
\usepackage{amssymb}

\begin{document}

\begin{center}
\begin{large}
\title\\{\textbf{ISGUR-WISE FUNCTION WITHIN A QCD QUARK MODEL WITH AIRY'S FUNCTION AS THE WAVE FUNCTION OF HEAVY-LIGHT MESONS}}\\\

\end{large}

\author\
\textbf{$Sabyasachi\;Roy^{\emph{1}}\footnotemark\;,  N\;S\;Bordoloi^{\emph{2}}\:and\:  D\:K\:Choudhury^{\emph{1,3,4}} $} \\\
\footnotetext{Corresponding author(On leave from Karimganj College,Assam, India). e-mail :  \emph{sroy.phys@gmail.com}}
\textbf{1}. Department of Physics, Gauhati University, Guwahati-781014, India.\\
\textbf{2}. Department of Physics, Cotton College, Guwahati-781001, India.\\
\textbf{3}. Centre for Theoretical Studies, Pandu College, Guwahati-781012,India.\\
\textbf{4}. Physics Academy of the North East, Guwahati-781014, India. \\

\begin{abstract}

We report some improved wave function for mesons taking linear confinement term in standard QCD potential as parent and Coulombic term as perturbation while applying quantum mechanical perturbation technique in solving Schr\"{o}dinger equation with such a potential. We find that  Airy's infinite series appears in the wave-function of the mesons. We report our calculations on the Isgur-Wise function and its derivatives for heavy-light mesons, within this framework.
\begin{flushleft}
\end{flushleft}
\end{abstract}
\end{center}

\section{Introduction:}\rm
\ \ \ Understanding the physics of hadrons, specially the heavy-light meson, has been the focus of interest in recent past. For heavy-light meson, the heavy-quark mass is much greater than $ \Lambda_{QCD} $, the QCD scale parameter. The four velocity of the heavy-quark is essentially the same as the four velocity of the meson and the heavy-quark, in the rest frame, appears as a static colour source. This gives rise to the consequence of spin-flavour symmetry in HQET and form factor suppression [1]. The overlap between the state of the light degrees of freedom in presence of a colour source moving with a velocity $ v $ , and that in the presence of a colour source moving with velocity $v^{\prime}$, is the Isgur-Wise function(IWF)[2]. This is the fundamental quantity in QCD, which can be determined non-perturbatively. \\
The motivation behind the calculation of IWF is to be able to extract CKM matrix elements $V_{cb}$ [3] from heavy-quark decays. Of particular interest are the semi-leptonic decays [4] like $\overline{B}\rightarrow Dl\overline{\nu}$ and $\overline{B}\rightarrow D^{*} l\overline{\nu}$ , from which it is possible to extract $V_{cb}$. \\
The main part of the IWF is the wave function of the meson [5] and some kinematic factor which depends upon four velocities of heavy-light mesons before and after recoil. Thus, the construction of wave function of meson is important for the detailed study of IWF  and its derivatives like slope (charge radii of meson) and curvature (convexity parameter).\\
For obtaining the meson wave function in potential model approach, the choice of potential is the first step. There are several potential models for quark-antiquark bound states like the Martin Potential[6], Cornell Potential[7], Richardson potential[8], Logarithmic potential[9]. The basic condition followed in enunciating these potentials are their flavour independence and existence of linear confinement. Of all the power law potentials, the Cornell potential is believed to be the more realistic phenomenological potential model for mesons as it is based upon extrapolation of two kinds of asymptotic - ultraviolet at short distance (Coulombic term) and infrared at large distance (linear term). It thus includes both the QCD concepts of `confinement' and `asymptotic freedom'. \\
 Starting from this Cornell potential of linear plus Coulombic form [10] and using two-body Schr\"{o}dinger equation, several successful attempts have been reported in recent years in extracting the wave function of heavy-light mesons . This has been done both with linear confinement term in potential as perturbation [11] and Coulombic term as perturbation [14]. \\
The case of Coulombic part in potential as parent in perturbation technique being simpler, several successful attempts have been made earlier [10-13] with such a choice, to extract meson wave function and make subsequent studies of meson properties. However, the origin of $q\overline{q}$ potential and also lattice QCD support the fact that the linear confinement term is the main contributing part to the potential. This inspires us to consider linear part in potential as parent and Coulombic part as perturbation. \\
The linear part in potential as parent leads to Airy's infinite polynomial function [16]  in unperturbed wave function. This brings some constraint in the perturbation technique in deducing the perturbed wave function with Coulombic part as perturbation. Some approximated form of the perturbed wave function has been reported recently [14] considering Airy function up to $ O(r^{3}) $ only, which obviously has its limitations. \\
In this paper, we report improvement of the previous wave function [14], and construct the meson wave function (with  Coulombic part in potential as perturbation) containing complete Airy's infinite series, following Dalgarno's perturbation method [17].\\
 Analysis of IWF with such a wave function containing Airy's infinite series, has been found to bring divergences. Very recently, we have overcome the problem of integrability  of the otherwise divergent infinite Airy's series in IWF, by successfully introducing some reasonable cut-off to the infinite upper limit of integration [15]. \\
In this paper also, we follow the same approach with our improved wave function and consider some cut-off to the infinite upper limit of integration while studying IWF and its derivatives. We calculate some reasonable range of this cut-off value which best suits our results, when compared with theoretical and experimental results for derivatives of IWF. We also study the sensitivity of the different orders of polynomial approximation of the infinite Airy's function in our calculation. \\
In section 2 we discuss the necessary formalism while section 3 contains the calculations and results. Summary and concluding remarks are stated in Section 4.

\section{Formalism:}
\subsection{Potential Model:}
The QCD inspired potential model of linear plus Coulombic type, has been reported in ref.[10-15]. For our studies of meson properties, we choose the widely accepted Cornell potential which has the form:
\begin{equation}
V (r) = -C_F \frac{\alpha_s}{r} + br + c
\end{equation}

$ C_F $ is the colour factor, which is given by :
\begin{equation}
C_F = \frac{N_C ^{2}-1}{2N_c}
\end{equation}
$ N_C $ is the colour quantum number; for $ N_C = 3 $, we have $ C_F = \frac{4}{3} $. \\
Here we take  $br$ as parent  so that our unperturbed Hamiltonian [18] is:
\begin{equation}
H_0 = -\frac{\nabla^2}{2\mu}+br
\end{equation}
with
\begin{equation}
H^\prime = -\frac{4\alpha_s}{3r}+ c
\end{equation}
as perturbation. Here $\mu$ is the reduced mass of the meson, which is given by:
\begin{equation}
   \mu= \frac{m_qm_Q}{m_q+m_Q}
\end{equation}
In the infinite quark mass limit ($m_Q\rightarrow\infty$), $ \mu \approx m_q $ .
We take $ b = 0.183 $ $GeV^2$ from charmonium spectroscopy [19] and $ c = 1 \; GeV $[20].
Under this consideration, the two body Schr\"{o}dinger equation [21,22] for the Hamiltonian $ H = H_0 + H^\prime$ is:
\begin{equation}
H|\Psi>=(H_0+H^\prime)|\Psi>=E|\Psi>
\end{equation}
\subsection{Wave Function:}
To find the unperturbed wave function corresponding to $H_0$, we use the two-body radial Schr\"{o}dinger equation (with $ \hbar =1 $):
\begin{equation}
[-\frac{1}{2\mu}(\frac{d^2}{dr^2}+\frac{2}{r}\frac{d}{dr})+br]R_0(r)=ER_0(r)
\end{equation}
where we have put $ V(r)=b r $. \\
For $ l=0 $ state, introducing $u(r) = r R_0 ( r )$ and dimensionless variable $\varrho$, we use the following formalism from ref. [23]:
\begin{equation}
\varrho=(2\mu b)^{1/3}r-(\frac{2\mu}{b^2})^{1/3}E
\end{equation}
The above equation (7) then reduces to:
\begin{equation}
\frac{d^2 u}{d\varrho^2}-\varrho u = 0
\end{equation}
The solution of this second order homogeneous differential equation [24] contains linear combination of two types of Airy's functions $Ai[r]$ and $Bi[r]$. From the nature of the Airy's function [25] we find that as $r\rightarrow \infty$, $Ai[r]\rightarrow 0$ and $Bi[r]\rightarrow \infty$. So, it is reasonable to reject the $Bi[r]$ part of the solution. The radial wave function thus has the form:
\begin{equation}
u(r)=NAi[(2\mu b)^{1/3} (r-\frac{E}{b})]
\end{equation}
where $N$ is our normalization constant which has the dimension of $GeV^{1/2}$. The boundary condition $u(0) = 0$ [26] gives us the unperturbed energy for ground state [23]:
\begin{equation}
W^{0}=E=-(\frac{b^2}{2\mu})^{1/3} \varrho_0
\end{equation}
where $\varrho_0$ is the zero of the Ai[r], such that Ai[$\varrho_0$]$=$0 [24] and it has the explicit form:
\begin{equation}
\varrho_0=-[\frac{3\pi(4n-1)}{8}]^{2/3}
\end{equation}
In our case we consider ground state wave function only ($n=1$). \\
The unperturbed wave function for ground state is thus obtained as:
\begin{equation}
\Psi^0(r)=\frac{N}{r}Ai[(2\mu b)^{1/3} r + \varrho_0]
\end{equation}
\begin{equation}
\Psi^0(r)=\frac{N}{r}Ai[\varrho_1 r + \varrho_0] = \frac{N}{2\sqrt{\pi}r}Ai[\varrho]
\end{equation}
where we have taken $\varrho_1$ $=$ $(2\mu b)^{1/3}$ and $ \varrho(r) = \varrho_1 r + \varrho_0 $. \\
The first order perturbed Eigen function $\Psi^{\prime}(r)$ and Eigen energy $W^{\prime}$ can be calculated using the following relation:
\begin{center}
\begin{equation}
H_0\Psi^\prime + H^\prime\Psi^0 = W^{0} \Psi^\prime + W^{\prime}\Psi^0
\end{equation}
\end{center}

We find,
\begin{center}
\begin{equation}
W^{\prime}=\int_0^\infty r^2H^\prime \mid\ \Psi^0 \mid^ 2 dr
\end{equation}
\end{center}
Now taking $ H^{\prime} = -\frac{B}{r}$ with $ B=\frac{4\alpha_s}{3}$, we obtain from equation  (15),
\begin{equation}
[-\frac{1}{2\mu}(\frac{d^2}{dr^2}+\frac{2}{r}\frac{d}{dr})+br-E]\Psi^{\prime}(r)=[\frac{B}{r}+W^{\prime}]\Psi^{0}(r)
\end{equation}
While applying Dalgarno's method of perturbation [27], we introduce the identity:
\begin{equation}
Ai ^{\prime}(\varrho)=\frac{dAi(\varrho)}{dr}=Z(\varrho)Ai(\varrho) \\
\end{equation}
so that
\begin{equation}
Ai ^{\prime\prime}(\varrho)= Z^{2}(\varrho)Ai(\varrho)+Z^{\prime}(\varrho)Ai(\varrho)
\end{equation}
Here, $ Z(\varrho)$ is a function of r, which can be calculated from the explicit polynomial expansion of Airy's function[Appendix-B]. \\
We obtain, after employing Dalgarno's perturbation method [Appendix-A],
\begin{equation}
\Psi^{\prime}(r)=\frac{N^{\prime}}{r}[A_1(r) r+A_2(r) r^{2}+A_3(r) r^{3}+ A_4(r) r^{4} + A_5(r) r^{5} +..... ]Ai[\varrho_1 r+\varrho_0]
\end{equation}
Here, $ A_1(r) , A_2(r), A_3(r) $ etc are obtained as :
\begin{eqnarray}
A_1(r)=-\frac{2\mu B}{2 \varrho_1 k_1 + \varrho_1 ^{2} k_2} \\
A_2(r)=-\frac{2\mu W^{\prime}}{2+4\varrho_1 k_1 + \varrho_1^{2} k_2}\\
A_3(r)= -\frac{2 \mu E A_1 }{6+6 \varrho_1 k_1 + \varrho_1^{2}k_2}  \\
A_4(r)= -\frac{2 \mu E A_2 - 2\mu b A_1}{12+ 8\varrho_1 k_1 + \varrho_1^{2}k_2} \\
A_5(r)= -\frac{2 \mu E A_3 - 2\mu b A_2}{20+ 10\varrho_1 k_1 + \varrho_1^{2}k_2} \\ \nonumber
\cdots\cdots\cdots\cdots \cdots\cdots\cdots\cdots \\ \nonumber
\cdots\cdots\cdots\cdots \cdots\cdots\cdots\cdots \\ \nonumber
\end{eqnarray}
where we take $ k_1(r)$ and $ k_2 (r)$ to be some function of r, as given below :
\begin{eqnarray}
Z(\varrho)=\frac{k_1(r)}{r} \\
Z^{2}(\varrho)+Z^{\prime}(\varrho)=\frac{k_2 (r)}{r^{2}}
\end{eqnarray}
Thus, we obtain the total wave function , as:
\begin{equation}
\Psi^{total}(r)=\frac{N^{\prime}}{r}[1+ {A_1(r) r+A_2(r) r^{2}+A_3(r) r^{3}+..... }]Ai[\varrho_1 r+\varrho_0]
\end{equation}
Here $N^\prime $ is the normalization constant of total wave function which also has the dimension of $GeV^{1/2}$.
Considering relativistic effect on the wave function, the total relativistic wave function is given by [21]:
\begin{equation}
\Psi_{rel}^{tot} (r) = \frac{N^{\prime}}{r}[1+ {A_1(r) r+A_2(r) r^{2}+A_3(r) r^{3}+..... }]Ai[\varrho_1 r+\varrho_0](\frac{r}{a_0})^{-\epsilon }
\end{equation}
Here,
\begin{equation}
a_0 = \frac{3 }{4\mu \alpha_s}= \frac{1}{B\mu} \;\; and\;\;  \epsilon = 1-\sqrt{1-(\frac{4\alpha_s}{3})^2}=1-\sqrt{1-B^2}
\end{equation}
The equations (20),(28) when compared with equations (16),(21) of ref[15], reveal the improvement of the present formalism over the previous one. Earlier, the perturbed wave-function was limited up to Airy function of order $r^3$, whereas in the present formalism, the perturbed wave function is the product of complete Airy's polynomial series and another infinite series.

\subsection{Isgur-Wise Function:}
 Under Heavy Quark Symmetry( HQS ), the strong interactions of the heavy quarks are independent of its spin and mass[28] and all the form factors are completely determined, at all momentum transfers, in terms of only one elastic form factor function, the universal Isgur-Wise function $\xi( v,v^\prime )$.  $\xi( v,v^\prime )$ depends only upon the four velocities   $v_\nu $ and $v_{\nu^\prime} $  of heavy particle before and after decay. This $\xi( v,v^\prime )$  is normalized at zero recoil [29].
If $y$ = $v_\nu $.$v_{\nu^\prime} $ , then, for zero recoil $(y=1)$, $\xi(y)=1$. With increasing recoil $y$ grows.
In explicit form IWF [30] can be expressed as:
\begin{equation}
\xi(y)=1-\rho^2 (y-1) +C(y-1)^2 + \cdots
\end{equation}
$\rho^2$ is the slope parameter at $y=1$, given by :
\begin{equation}
\rho^2 = -\frac{\delta\xi (y)}{\delta y}|_{y=1}
\end{equation}
\begin{flushright}
$\rho$ is known as the charge radius. \\
\end{flushright}
C is the convexity parameter given by :
\begin{equation}
C= \frac{\delta^{2}\xi (y)}{\delta y^2}|_{y=1}
\end{equation}
The calculation of this IWF is non-perturbative in principle and its depends upon the meson wave function and some kinematic factor, as given below :
\begin{equation}
\xi(y)=\int_0 ^\infty 4\pi r^2 |\Psi(r)|^2\cos(pr)dr
\end{equation}
where $\cos(pr)=1-\frac{p^2 r^2}{2}+\frac{p^4 r^4}{24}$ +$\cdot\cdot\cdot\cdot\cdot\cdot$  with $ p^2=2\mu^2 (y-1)$. Taking $cos(pr)$ up to  $O(r^4)$ we get from equation (34),

\begin{flushleft}
\begin{eqnarray}
\xi(y)= \int_0 ^\infty 4\pi r^2 |\Psi(r)|^2dr - [4\pi\mu^2\int_0^\infty r^4|\Psi(r)|^2dr](y-1)+[\frac{2}{3}\pi\mu^4\int_0^\infty r^6|\Psi(r)|^2dr](y-1)^2
\end{eqnarray}
\end{flushleft}

Equations (31) and (35) give us :
\begin{eqnarray}
\rho^2 = 4\pi\mu^2\int_0^\infty r^4|\Psi(r)|^2dr \\
C= \frac{2}{3}\pi\mu^4\int_0^\infty r^6|\Psi(r)|^2dr \;\;\;\; and \\
\int_0 ^\infty 4\pi r^2 |\Psi(r)|^2dr =1
\end{eqnarray}
Equation (38) gives the normalization constants $ N \;\;$ and $\;\;  N^{\prime}$ for $\Psi^0 (r)$ and $\Psi_{tot}^{rel} (r)$  .
\section{Calculation and result:}
From equation (29), we find that the total wave function contains two infinite series - one a power series in $ r $ with coefficients $ A_1(r), A_2(r) $ etc and the other an infinite Airy's series (equation(B.1) in appendix-B). As the infinite limit of integration in IWF brings divergence of both the series, we are compelled to consider some reasonable cut-off  $r_0$ to this upper limit of integration. In principle, this cut-off $r_0$ should be greater than the size of hadrons. In our calculations, we consider this $r_0$ to be greater than $\frac{1}{m_h}$, $m_h$ being the mass of hadron. This always keeps $r_0$ to be much greater than size of meson concerned. Also,consideration of such cut-off to upper limit of integrations will not sacrifice the nature and value of IWF and its derivatives, because, Airy's function falls very sharply (almost exponentially) and almost dies out with increasing r-value [31]. In fact, the Airy's function value becomes negligibly small for $r > 4 $ ( $AiryAi[4]=0.000952 $).\\
Also, the graph of normalization constant ($N^\prime$ )versus the cut-off to upper limit ( $r_0$ ) shows that $N^{\prime}$ value decreases with increase in $r_0$. For D meson, as a representative case, such variation is shown in fig 1. \\

$k_1 $ and $k_2 $ in equations (21-25) are calculated considering lowest Airy polynomial order, as [Appendix-B]:
\begin{eqnarray}
k_1(r)=1+\frac{k}{r}\\
k_2(r)=\frac{k^2}{r^2}\\
with  \;\;\; k=\frac{a_1-b_1 \varrho_0}{b_1 \varrho_1}
\end{eqnarray}
Here, $a_1$ and $b_1$ are constants in Airy's infinite polynomial  with $a_1=\frac{1}{3^{2/3} \Gamma(2/3)}=0.3550281$ and $ b_1=\frac{1}{3^{1/3} \Gamma(1/3)} =0.2588194$. \\

$A_1(r)$ and $A_2(r)$ etc as in equations(21-25) then have more explicit forms like:
\begin{eqnarray}
A_1(r)=\frac{-2 \mu B r^{2}}{2\varrho_1 r^{2}+2 \varrho_1 k r + \varrho_1^{2} k^{2}} \\
A_2(r)=\frac{-2\mu W^{\prime}r^{2}}{(2+4\varrho_1)r^{2}+4\varrho_1 k r+ \varrho_1^{2}k^{2}}\;\; etc. \\ \nonumber
\end{eqnarray}
With these more explicit functions of $r$ in the meson wave function as in equation (29), we have explored  $\xi(y)$  and its derivatives for Airy polynomial order up to $ r^{10}$, taking different cut-off values ranging from $r_0=3 \; GeV^{-1}$ to $r_0 = 7 \;GeV^{-1}$, for $ D, D_s, B, B_s $ mesons taking  $\alpha_s=0.22$ [32]. The results are shown in Tables 1 to 4. \\
\begin{table}[tb]
\begin{center}
\caption{$\rho^{2}$ and $C$ for D meson ( $ \mu=0.2761$)}\label{centre}
\begin{tabular}{|c|ccc|}
  \hline
      $r_0 \;(in\;GeV^{-1})$ & $N^{\prime}$ & $\rho^2$ & C  \\
   \hline
   5.0    & 0.6616 & 0.6821 & 0.1138 \\    \hline
   5.5  & 0.5249 & 1.2249 & 0.3571 \\    \hline
   6.0    & 0.3655 & 1.9099 & 0.7342 \\    \hline
   6.5  & 0.2566 & 2.463  & 1.1247 \\    \hline

\end{tabular}
\end{center}
\end{table}
\begin{table}[tb]
\begin{center}
\caption{$\rho^{2}$ and $C $ for B meson ( $ \mu=0.318$)}\label{centre}
\begin{tabular}{|c|ccc|}
  \hline
      $r_0 \;(in\;GeV^{-1})$ & $N^{\prime}$ & $\rho^2$ & C  \\
   \hline
   4.0      & 0.7277 & 0.6699   & 0.0979 \\    \hline
   4.5      & 0.7184 & 0.7030   & 0.1118 \\    \hline
   5.0      & 0.6069 & 1.1758   & 0.3451 \\    \hline
   5.5      & 0.4172 & 2.0523   & 0.8722 \\    \hline
   6.0      & 0.2792 & 2.7854   & 1.4441 \\    \hline

\end{tabular}
\end{center}
\end{table}

\begin{table}[tb]
\begin{center}
\caption{$\rho^{2}$ and $C $ for $D_s$ meson ( $ \mu=0.368 $)}\label{centre}
\begin{tabular}{|c|ccc|}
  \hline
      $r_0 \;(in\;GeV^{-1})$ & $N^{\prime}$ & $\rho^2$ & C  \\
   \hline
   4.4      & 0.7648 & 0.7968 & 0.1388 \\    \hline
   5.5      & 0.7014 & 1.0707 & 0.2882 \\    \hline
   5.0      & 0.4947 & 2.1054 & 0.9665 \\    \hline
   5.5      & 0.3145 & 3.1059 & 1.8176 \\    \hline
   6.0      & 0.2121 & 3.8782 & 2.6875 \\    \hline
\end{tabular}
\end{center}
\end{table}

\begin{table}[tb]
\begin{center}
\caption{$\rho^{2}$ and $C $ for $B_s$ meson ( $ \mu=0.4401 $)}\label{centre}
\begin{tabular}{|c|ccc|}
  \hline
      $r_0 (in\;GeV^{-1})$ & $N^{\prime}$ & $\rho^2$ & C  \\
   \hline
   3.0      & 0.8511 & 0.8792   & 0.1651 \\    \hline
   3.5      & 0.8087 & 0.9809   & 0.2101 \\    \hline
   4.0      & 0.7855 & 1.0924   & 0.2805 \\    \hline
   4.5      & 0.5825 & 2.2312   & 1.1423 \\    \hline
   5.0      & 0.3512 & 3.6471   & 2.5282 \\    \hline
\end{tabular}
\end{center}
\end{table}

We find that, for the given Airy order, with increase in cut-off value, $\rho^2$ and $C$ values increase steadily. However, the results show closer resemblance to recent results [33-41]( as shown in Table 5) up to a specific cut-off value $r_0$ for different mesons. For such specific range and order, our results show improvement over the result of ref [8]. At cut-off value higher than $r_0 \sim 6 \;GeV^{-1}$, the results jump to higher values than our expectations ( Figure 2(a) and 2(b)).\\
\begin{table}[tb]
\begin{center}
\caption{Results of slope and curvature of $\xi(y$) in different models and collaborations.}\label{cross}
\begin{tabular}{|c|r|r|}
  \hline
Model / collaboration &	Value of slope & Value of curvature \\
\hline \hline
 Ref [14] & 0.7936 & 0.0008 \\
 Le Youanc et al [33] & $\geq 0.75$ & $\geq 0.47$ \\
 Skryme Model [34] & 1.3 & 0.85 \\
 Neubert [35] & 0.82$\pm$0.09 & -- \\
 UK QCD Collab. [36]  & 0.83 & -- \\
 CLEO [37,38] & 1.67 & -- \\
 BELLE  [39] & 1.35 & -- \\
HFAG [40] & 1.17 $\pm 0.05$ & -- \\
Huang [41] & 1.35 $\pm 0.12$ & -- \\

  \hline
\end{tabular}
\end{center}
\end{table}

Regarding sensitivity of the order of polynomial in infinite Airy's function, the results for $\rho^2$ and $C$ do not differ much upon variation of order of polynomial in Airy's function from $ r^{4} $ to $r^{10} $. The results of such study in the case of D meson is shown in Table 6. \\
\begin{table}[tb]
\begin{center}
\caption{Sensitivity of the polynomial order of Airy's function( for D meson with $r_0=5.48$)}\label{centre}
\begin{tabular}{|c|c|c|}   \hline
      Airy Order &  $\rho^2$ & C  \\
   \hline
   $r^{4}$      & 1.6338 & 0.3345   \\    \hline
   $r^{6}$     & 1.1657 & 0.3346    \\    \hline
   $r^{7}$     & 1.1666 & 0.3347   \\    \hline
   $r^{9}$     & 1.1665 & 0.33467    \\    \hline
   $r^{10}$   & 1.16644 & 0.334674    \\    \hline
    With complete Airy series \\(After numerical integration)  & 1.16644 & 0.334674    \\    \hline
\end{tabular}
\end{center}
\end{table}

In all calculations, we have taken care that the boundary condition of IWF ( $\xi(1)=1$ ) is maintained throughout. The variation of $\xi(y)$ with y for different cut-off value $r_0$ ,with Airy order $r^{10}$, for different mesons is shown in Figures 3(a) to 3(d).

\section{Conclusion and remarks:}
In this work, we have deduced the wave function for mesons involving the complete Airy's polynomial series, without introducing any approximation. The wave function constructed here is certainly an improvement over that in references [14,15]- in which the perturbed wave function $\Psi^{\prime}(r)$  has been calculated using Airy order $r^3$. Our  perturbed wave function(and hence total wave function) contains complete Airy's series.\\
In our earlier work [15], we have shown that cutting off the upper limit of integrations in $\xi(y)$ and its derivatives to some reasonable point does not upset the result, rather it almost conforms to the experimental expectations. In this work, we have followed similar approach in the study of IWF. Such study also, in turn, confirms the compatibility of our formalism.\\
Also, for each value of cut-off $r_0$ , we have considered the asymptotic form of the Airy's function taking limits of integration from $r_0$  to $\infty$ .
\begin{equation}
Ai[\varrho]_{asympt} \sim \frac{ \exp{(-\frac{2}{3}\varrho^{3/2}})}{2\sqrt{\pi}\varrho^{1/4}}
\end{equation}
With this asymptotic form we have also calculated the derivatives of $\xi(y)$. Such analysis shows that, taking this asymptotic form of Airy's function very small values of  $\rho^2$ and C are obtained(Table 7). This, thus, also confirms that the margin of error, by considering cut-off to infinite upper limit of integration, is negligible.\\

\begin{table}[tb]
\begin{center}
\caption{Values of  $\rho^2$ and C with asymptotic form of Airy's function.}\label{centre}
\begin{tabular}{|c|c|c|}
  \hline
  $r_0 \;(in\;GeV^{-1})$  & $\rho^2$ (asymptotic ) & C(asymptotic)\\
  \hline \hline
   3 & $8.4\times  10^{-6}$ & $1.2\times  10^{-6}$  \\
   \hline
   4 & $2.6\times  10^{-7}$ & $6.0\times  10^{-8}$ \\
   \hline
   5 & $4.6 \times  10^{-9}$ & $1.6 \times 10^{-9}$ \\
   \hline
   6 & $5.027 \times  10^{-11}$ &	$2.464 \times 10^{-11}$ \\
   \hline
   7& $3.56 \times  10^{-13}$ &	$2.345 \times 10^{-13}$ \\
   \hline
   8&  $1.695 \times  10^{-15}$ 	& $ 7.028 \times 10^{-15}$ \\
   \hline
  \end{tabular}
\end{center}
\end{table}
 The figures 3(a) to 3(d) confirm the fact that boundary condition for zero recoil $(\xi(1)=1)$ is maintained all through, with given polynomial orders of Airy's function and for different cut-off values. \\
Regarding the results of ref [14] , which is for Airy order $r^{3}$ , we find that results match with our cut-off value $r_0 = 5.095 \; GeV^{-1}$ for D meson. \\
Further, we study the compatibility of our potential model with the recent results of Heavy Flavour Averaging Group(HFAG) [40]. Taking the result of $\rho^2$ , we fix the value of cut-off for different orders of polynomial in Airy function (Table 8). This indicates that, the range of cut-off  value $r_0 = 4 \; GeV^{-1}$ to $r_0 = 6 \;  GeV^{-1}$  matches the expectations of ref [40]. \\\\
\begin{table}[tb]
\begin{center}
\caption{Value of cut-off $r_0$ fixing $\rho^{2}=1.17$ from ref[40]}\label{centre}
\begin{tabular}{|c|c|}
  \hline
      $meson$ & $r_0 (in\; GeV^{-1})$  \\    \hline
   D        & 5.48  \\    \hline
   B        & 4.998  \\   \hline
   $D_s$    & 4.54  \\    \hline
   $B_s$    & 4.05  \\    \hline
\end{tabular}
\end{center}
\end{table}

W conclude by making the following comments on our present approach.
\begin{itemize}
  \item We opt for the quantum mechanical perturbation technique in deducing the meson wave function, due to the constraint in getting exact analytic solution of Schrodinger equation involving linear plus Coulombic type potential.
  \item It is well known that method of numerical solution of Schr\"{o}dinger equation [42] or Lattice QCD [43] gives more accurate results compared to relatively crude potential model approach following perturbation technique. However, the potential model approach is believed to give more physical insight into the problem under consideration.
  \item Lastly, in our analysis of IWF, the parameters $k_1(r)$ and $k_2(r)$ and hence $A_1(r), A_2(r)$ etc are calculated considering lowest order in Airy polynomial. Considering higher Airy polynomial order in such calculations may bring further refinement of the result.
\end{itemize}

\paragraph{Acknowledgement :\\ }
\begin{flushleft}
\emph{One of the authors (SR) acknowledges the support of University Grants Commission, Govt. of India in terms of fellowship under Faculty Development Programme, to pursue research work at Gauhati University.}
\end{flushleft}

\appendix
\numberwithin{equation}{section}
\begin{center}
\section{Appendix}
\end{center}

From equation (17), in terms of radial wave function $ R(r)$ we get:
\begin{equation}
[-\frac{1}{2\mu}(\frac{d^2}{dr^2}+\frac{2}{r}\frac{d}{dr})+b r-E]R(r)=[\frac{B}{r}+W^{\prime}]\frac{1}{r}Ai[\varrho]
\end{equation}
Let,
\begin{equation}
R(r)=\frac{1}{r}F(r)Ai[\varrho]= \frac{1}{r}F(r)Ai[\varrho_1 r + \varrho_0]
\end{equation}
So that,
\begin{equation}
\frac{dR}{dr}=-\frac{1}{r^{2}}F(r)Ai[\varrho]+\frac{1}{r}F^{\prime}(r)Ai[\varrho]+\frac{\varrho_1}{r}F(r)Ai^{\prime}[\varrho]
\end{equation}
and
\begin{equation}
\frac{d^{2}R}{dr^{2}}=\frac{2}{r^{3}}F(r)Ai[\varrho]-\frac{2}{r^{2}}F^{\prime}(r)Ai[\varrho]-\frac{2\varrho_1}{r^{2}}F(r)Ai^{\prime}[\varrho]+\frac{1}{r}F^{\prime\prime}(r)Ai[\varrho]\\ +\frac{2\varrho_1}{r}F^{\prime}(r)Ai^{\prime}[\varrho]+\frac{\varrho_1^{2}}{r}F(r)Ai^{\prime\prime}[\varrho]
\end{equation}
Now we introduce the identity :
\begin{equation}
Ai^{\prime}[\varrho]=\frac{d Ai(\varrho)}{dr}=Z(\varrho)Ai(\varrho)
\end{equation}
so that
\begin{equation}
Ai ^{\prime\prime}(\varrho)= Z^{2}(\varrho)Ai(\varrho)+Z^{\prime}(\varrho)Ai(\varrho)
\end{equation}
Then the equation (A.1) becomes :
\begin{equation}
\frac{1}{r}F^{\prime\prime}(r)+\frac{2\varrho_1}{r}F^{\prime}(r)Z(\varrho)+\frac{\varrho_1^{2}}{r}F(r)Z^{\prime}(r)
+\frac{\varrho_1^{2}}{r}F(r)Z^{2}(r)-2\mu(b r -E)\frac{1}{r}F(r)=  \\
-2\mu (\frac{B}{r^{2}}+\frac{W^{\prime}}{r})
\end{equation}
so that,
\begin{equation}
F^{\prime\prime}(r)+2\varrho_1 F^{\prime}(r)Z(\varrho)+ \varrho_1^{2}[Z^{2}(\varrho)+Z^{\prime}(\varrho)]F(r)-2\mu(b r-E)F(r)= \\
-\frac{2\mu B}{r}-2\mu W^{\prime}
\end{equation}
Assuming
\begin{eqnarray}
Z(\varrho)=\frac{k_1(r)}{r} \\
Z^{2}(\varrho)+Z^{\prime}(\varrho)=\frac{k_2(r)}{r^{2}}
\end{eqnarray}
We get,
\begin{equation}
 F^{\prime\prime}(r)+2\varrho_1 F^{\prime}(r)\frac{k_1(r)}{r}+\varrho_1^{2} F(r)\frac{k_2}{r^{2}}-2\mu(b r-E)F(r)=-\frac{2\mu B}{r}-2\mu W^{\prime}
\end{equation}

We take,
\begin{eqnarray}
F(r)=\sum_l A_l r^{l} \\
F^{\prime}(r)=l\sum_l A_l r^{l-1} \\
F^{\prime\prime}(r)=l(l-1)\sum_l A_l r^{l-2}
\end{eqnarray}
so that equation (A.11) becomes,

\begin{equation}
l(l-1)\sum_l A_l r^{l-2} +2\varrho_1k_1 l \sum_l A_l r^{l-2} +\varrho_1^{2} k_2 \sum_l A_l r^{l-2}+2\mu E\sum_l A_l r^{l}-2\mu b\sum_l A_l r^{l+1}=\frac{2\mu B}{r}-2\mu W^{\prime}
\end{equation}

Equating powers of $ r^{-2}$ on both sides of equation (A.15) we get :
\begin{equation}
\varrho_1^{2}k_2 A_0 =0
\end{equation}
This gives $ A_0 =0 $.\\
Further, equating powers of $ r^{-1}, r^{0}, r^{1}, r^{2}$ and $r^{3}$ of both sides of equation (A.15), we get:
\begin{eqnarray}
2\varrho_1 k_1 A_1 + \varrho_1^{2} k_2 A_1=-2\mu B \\
2A_2 +4\varrho_1k_1 A_2 +\varrho_1^{2} k_2 A_2=-2\mu W^{\prime} \\
6A_3 + 6\varrho_1k_1A_3 +\varrho_1^{2} k_2 A_3 +2\mu E A_1=0 \\
12A_4+8\varrho_1 k_1 A_4+\varrho_1^{2}k_2 A_4 +2\mu EA_2-2\mu B A_1=0 \\
20A_5+10\varrho_1 k_1A_5 +\varrho_1^{2}k_2A_5+2\mu EA_3-2\mu b A_2 =0
\end{eqnarray}
From these equations (A.17-A.21), we extract the expressions of $ A_1(r), A_2(r), A_3(r), A_4(r)$ and $ A_5(r) $ as:
\begin{flushleft}
\begin{eqnarray}
A_1(r)=-\frac{2\mu B}{2 \varrho_1 k_1 + \varrho_1 ^{2} k_2} \\
A_2(r)=-\frac{2\mu W^{\prime}}{2+4\varrho_1 k_1 + \varrho_1^{2} k_2}\\
A_3(r)= -\frac{2 \mu E A_1 }{6+6 \varrho_1 k_1 + \varrho_1^{2}k_2}  \\
A_4(r)= -\frac{2 \mu E A_2 - 2\mu b A_1}{12+ 8\varrho_1 k_1 + \varrho_1^{2}k_2} \\
A_5(r)= -\frac{2 \mu E A_3 - 2\mu b A_2}{20+ 10\varrho_1 k_1 + \varrho_1^{2}k_2} \\
\end{eqnarray}
\end{flushleft}

\begin{equation}
F(r)=\sum_l A_l r^{l} = A_1(r) r + A_2(r) r^{2} + A_3(r) r^{3} + A_4(r) r^{4} + A_5(r) r^{5} +\cdots\cdots\cdots\cdots
\end{equation}
Ultimately , we get the perturbed wave-function $ \Psi^{\prime}(r) $ as:
\begin{equation}
\Psi^{\prime}(r)=\frac{N^{\prime}}{r}[A_1(r) r+A_2(r) r^{2}+A_3(r) r^{3}+ A_4(r) r^{4} + A_5(r) r^{5} +..... ]Ai(\varrho_1 r+\varrho_0)
\end{equation}
\\

\numberwithin{equation}{section}
\begin{center}
\section{Appendix}
\end{center}

The Airy's infinite series as a function of $\varrho =\varrho_1 r +\varrho_0 $ can be expressed as [24] :
\begin{eqnarray}
Ai[\varrho_1 r + \varrho_0] = a_1[1+\frac{(\varrho_1 r + \varrho_0)^3}{6}+\frac{(\varrho_1 r + \varrho_0)^6}{180}+\frac{(\varrho_1 r + \varrho_0)^9}{12960}+...]- \nonumber \\
 b_1[(\varrho_1 r + \varrho_0) +\frac{(\varrho_1 r + \varrho_0)^4}{12}+\frac{(\varrho_1 r + \varrho_0)^7}{504}+\frac{(\varrho_1 r + \varrho_0)10}{45360}+...]
\end{eqnarray}

To find $ k_1(r)$ and $k_2(r) $ , we take truncated Airy series up to lowest order, so that, we have :
\begin{eqnarray}
Z(r)=\frac{k_1(r)}{r} = \frac{Ai^{\prime}(\varrho_1 r+\varrho_0)}{Ai(\varrho_1 r+\varrho_0)} \\
=\frac{-b_1 \varrho_1}{a_1-b_1(\varrho_1 r+\varrho_0)} \\
=\frac{b_1 \varrho_1}{b_1(\varrho_1 r+\varrho_0)-a_1} \\
=\frac{1}{r}[1-\frac{a_1-b_1\varrho_0}{b_1\varrho_1}\frac{1}{r}]^{-1} \\
=\frac{1}{r}[1-\frac{k}{r}]^{-1}
=\frac{1}{r}(1+\frac{k}{r}) \\ \nonumber
\end{eqnarray}
Therefore,
\begin{equation}
k_1(r)=1+\frac{k}{r}
\end{equation}
and
\begin{equation}
Z(r)=\frac{1}{r}+\frac{k}{r^2}
\end{equation}
with
\begin{equation}
k=\frac{a_1-b_1 \varrho_0}{b_1 \varrho_1}
\end{equation}
Also,

\begin{equation}
Z^{2}(r)=\frac{1}{r^{2}}(1+\frac{2k}{r}+\frac{k^{2}}{r^{2}})=\frac{1}{r^{2}}+\frac{2k}{r^{3}}+\frac{k^{2}}{r^{4}}
\end{equation}
and
\begin{equation}
Z^{\prime}(r)=-\frac{1}{r^{2}}-\frac{2k}{r^{3}}
\end{equation}
so that,
\begin{equation}
\frac{k_2(r)}{r^{2}}=Z^{2}(r)+Z^{\prime}(r)=\frac{k^{2}}{r^{4}}
\end{equation}
We thus obtain,
\begin{equation}
k_2(r)=\frac{k^{2}}{r^{2}}
\end{equation}
\\
\\
\\

\begin{figure}[h]
    \centering
    \includegraphics{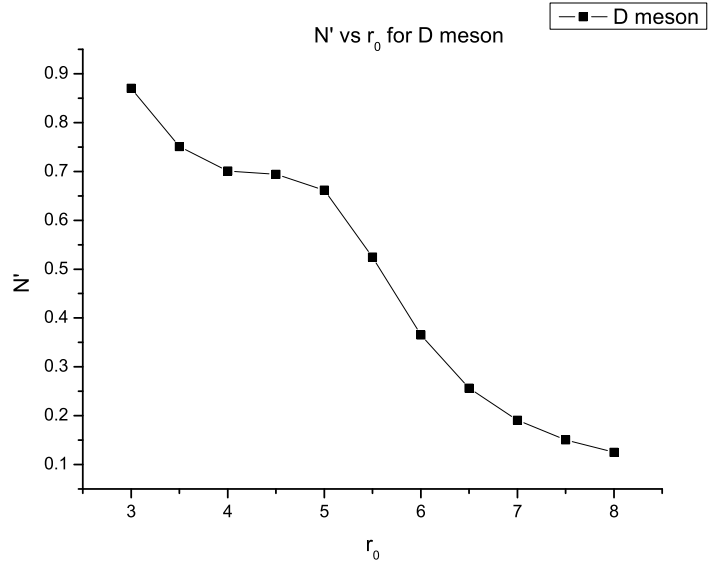}
    \caption{ Variation of $N^{\prime}$ with $r_0$}
    \label{Ai[r] vs r}
\end{figure}

\begin{figure}
    \centering
    \subfigure[ $\rho^{2}$ vs $r_0$]
    {
        \includegraphics[width=3.0in]{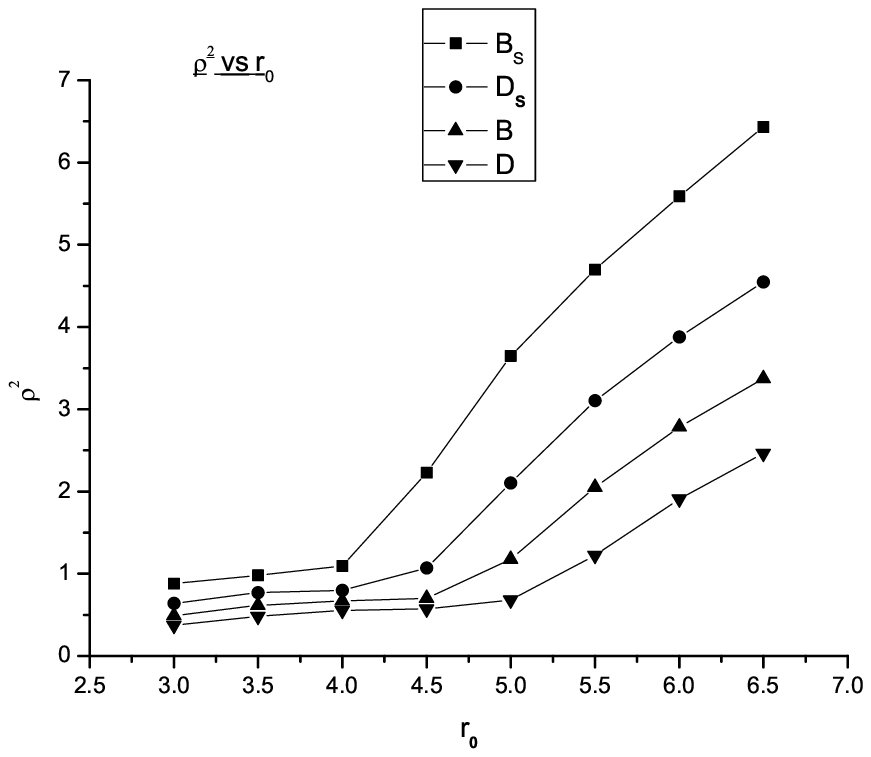}
        \label{fig:first_sub}
    }
        \subfigure[ $C$ vs $r_0$]
    {
        \includegraphics[width=3.0in]{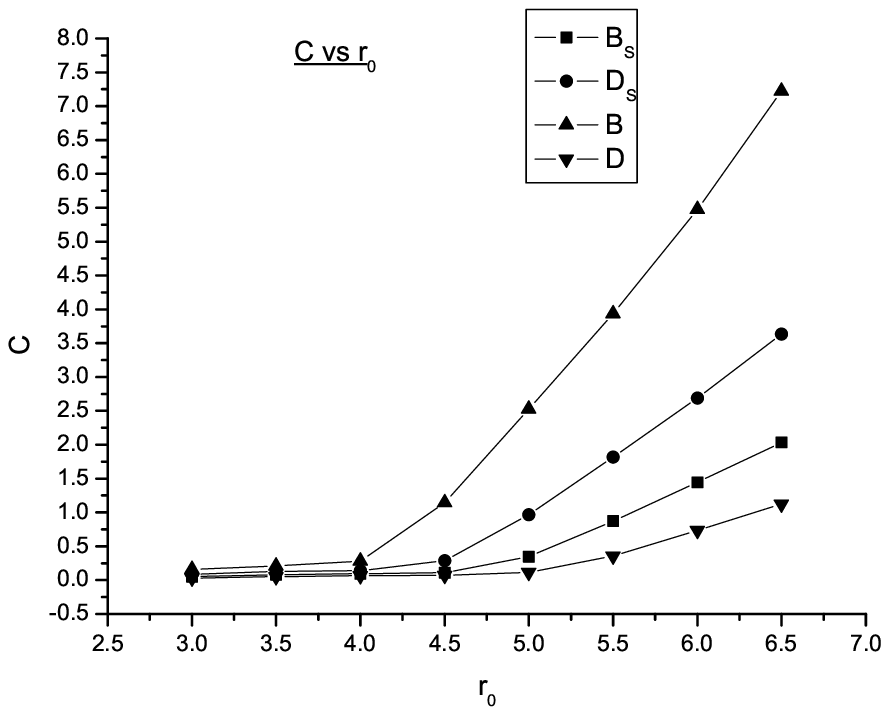}
        \label{fig:second_sub}
    }
    \caption{Variation of $\rho^{2}$ and $C$ with $r_0$ for different mesons}
    \label{fig:sample_subfigures}
\end{figure}

\begin{figure}
    \centering
    \subfigure[ $\xi(y)$ vs $ y$ for D meson]
    {
        \includegraphics[width=3.0in]{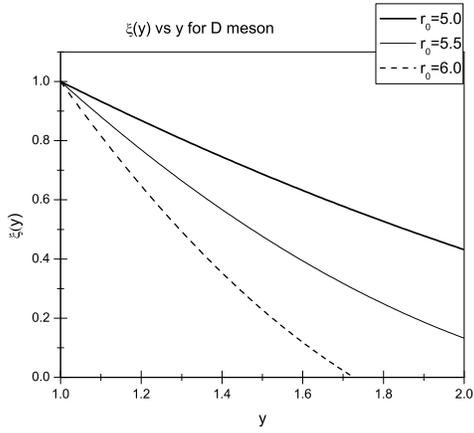}
        \label{fig:first_sub}
    }
        \subfigure[ $\xi(y)$ vs $ y$ for B meson]
    {
        \includegraphics[width=3.0in]{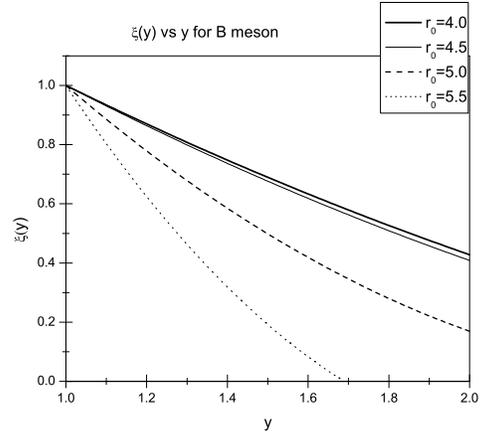}
        \label{fig:second_sub}
    }
    \\
    \subfigure[ $\xi(y)$ vs $ y$ for $D_s$ meson]
    {
        \includegraphics[width=3.0in]{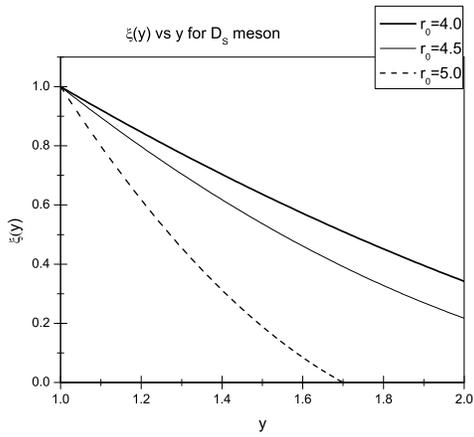}
        \label{fig:third_sub}
    }
        \subfigure[ $\xi(y)$ vs $ y$ for $B_s$ meson]
    {
        \includegraphics[width=3.0in]{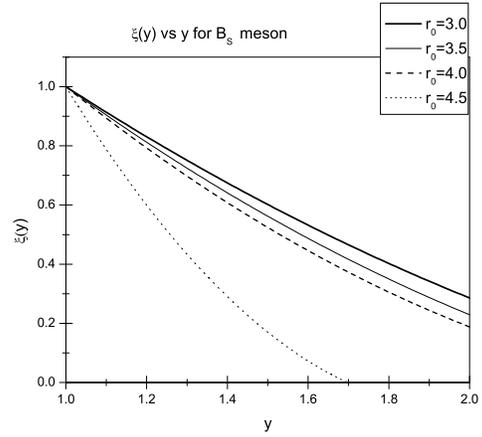}
        \label{fig:fourth_sub}
    }

    \caption{Variation of $\xi(y)$ with $ y$ for different mesons}
    \label{fig:sample_subfigures}
\end{figure}

\begin{center}
-----
\end{center}

\end{document}